\documentclass[10pt, conference, letterpaper]{IEEEtran}
\IEEEoverridecommandlockouts
\usepackage{cite}
\usepackage{amsmath,amssymb,amsfonts}
\usepackage{algorithmic}
\usepackage{graphicx}
\usepackage{textcomp}
\usepackage{xcolor}
\usepackage[normalem]{ulem}
\usepackage[colorlinks=true,citecolor=green,urlcolor=blue,bookmarks=false,hypertexnames=true]{hyperref} 
\usepackage{multirow}
\usepackage{subfigure}

\def\BibTeX{{\rm B\kern-.05em{\sc i\kern-.025em b}\kern-.08em
    T\kern-.1667em\lower.7ex\hbox{E}\kern-.125emX}}
\begin{document}


\title{Social Distancing Alert with Smartwatches}
\author{\IEEEauthorblockN{Xin Wang$^{1*}$, Xilei Wu$^{1*}$, Huina Meng$^{1}$, Yuhan Fan$^2$, Jingang Shi$^1$, Han Ding$^1$, Fei Wang$^{1\#}$
}
\thanks{*equal contribution, \#corresponding author.}
\IEEEauthorblockA{\textit{1 Xi'an Jiaotong University} Xi'an Shaanxi, China, 710049 \\
\textit{2 Harbin Institute of Technology}, Harbin Heilongjiang, China, 150001 \\
\{xwang6, xlwuuu, menghuina\}@stu.xjtu.edu.cn, 1171910109@stu.hit.edu.cn, \{jingang, dinghanxjtu, feynmanw\}@xjtu.edu.cn}
}

\maketitle

\begin{abstract}
Social distancing is an efficient public health practice during the COVID-19 pandemic. However, people would violate the social distancing practice unconsciously when they conduct some social activities such as handshaking, hugging, kissing on the face or forehead, etc. In this paper, we present SoDA, a social distancing practice violation alert system based on smartwatches, for preventing COVID-19 virus transmission. SoDA utilizes recordings of accelerometers and gyroscopes to recognize activities that may violate social distancing practice with simple yet effective Vision Transformer models. Extensive experiments over 10 volunteers and 1800+ samples demonstrate that SoDA achieves social activity recognition with the accuracy of 94.7\%, 1.8\% negative alert, and 2.2\% missing alert. ~\textit{Code 
is publicly available at \url{https://github.com/aiotgroup/SoDA}}.

\end{abstract}



\section{Introduction}~\label{sec:introduction}
Social distancing, maintaining approximately 6 feet or 2 meters from others, is an efficient public health practice during the COVID-19 pandemic that aims to prevent people who are infected from coming in close contact with healthy people in order to reduce virus transmission. However, people would unconsciously violate the social distancing practice when conducting some daily social activities such as shaking hands, hugging, kissing on the face or forehead, etc. To reduce this unconscious violation, we propose SoDA which leverages accelerometers and gyroscopes of smartwatches to recognize these social activities, serving as an alert system for promoting users' adherence to the social distancing practice.

There are flourishing works on social distance estimation after the outbreak of the COVID-19\cite{ghasemi2021auto,chandel2020proxitrak,kanjo2021crowdtracing,reddy2020social,li2021smartdistance,bian2020wearable}. For example, surveillance cameras are utilized to estimate the density of people for evaluating their adherence to the social distancing practice~\cite{ghasemi2021auto}. 
Some wireless communication technologies, such as Bluetooth Low Energy~\cite{chandel2020proxitrak}, Wi-Fi~\cite{kanjo2021crowdtracing}, and Ultra-Wide Band~\cite{reddy2020social}, are adopted to localize people indoors. Besides, acoustic sensors~\cite{li2021smartdistance} and magnetic sensors~\cite{bian2020wearable} are to estimate the distance between users who carry the sensory system. Still, we propose SoDA and highlight our motivations below.
\begin{itemize}
    \item Systems based on cameras or wireless communication technologies are proposed for business users, such as malls, schools, or metro stations. An individual person cannot receive a specific alert efficiently if she/he violates the social distancing practice. SoDA works along with users and can alert users specifically in time.
    
    \item Systems based on acoustic sensors and magnetic sensors only estimate the distance between people who carry the systems simultaneously. The strict cooperation seriously harms the usage. SoDA is only required to be worn by one user, and report the alert for activities that would violate the social distancing practice.
\end{itemize}

\begin{figure}[t]
    \centering
    \includegraphics[width=0.98\linewidth]{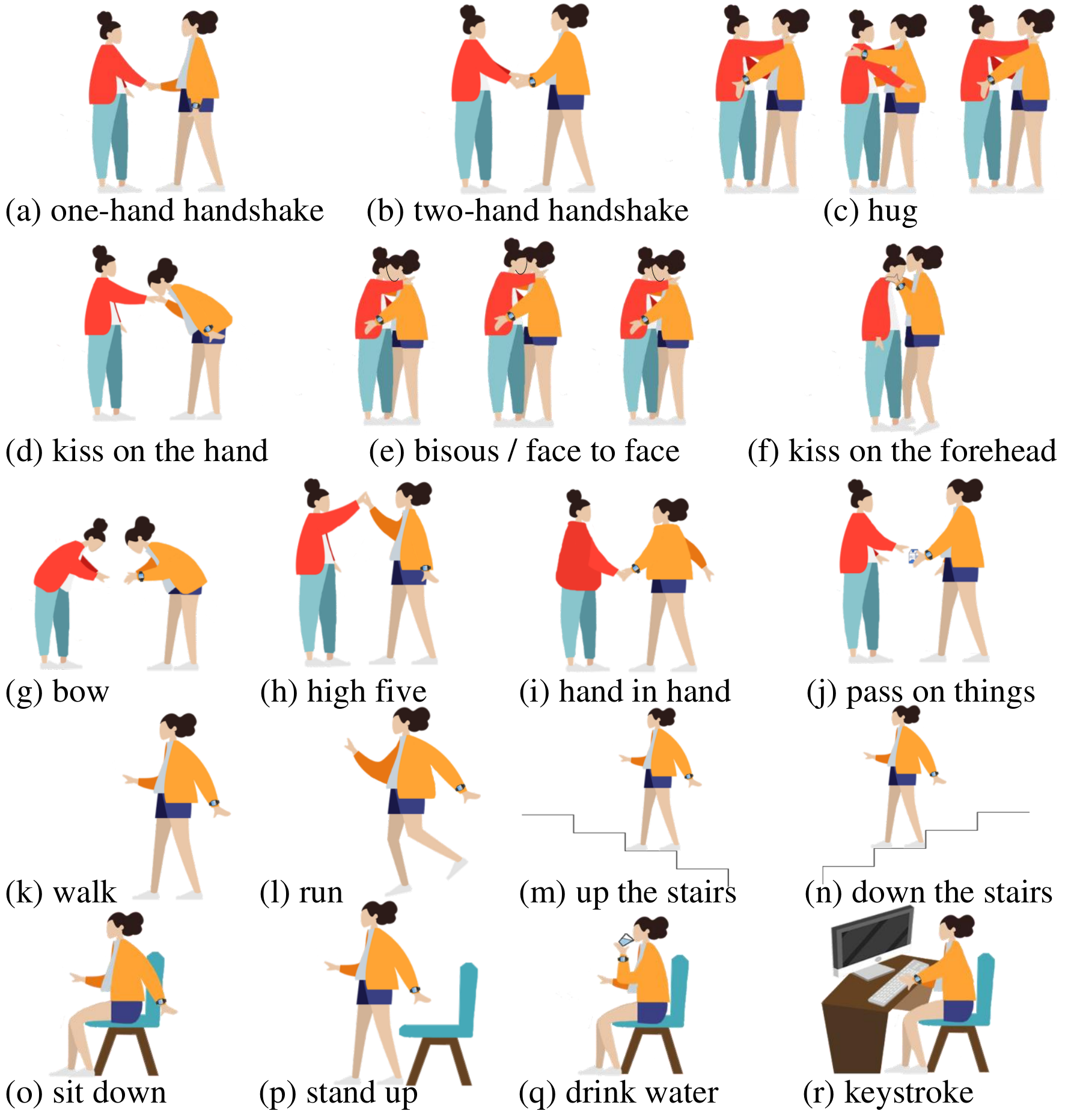}
    \caption{Some social activities, e.g. (a-j), naturally violate the social distancing practice during the COVID-19 pandemic. We propose SoDA, an intelligent approach based on smartwatches, to distinguish these ten social activities from other eight daily activities, i.e. (k-r), to serve as a violation alert for promoting users' adherence to the social distancing practice, reducing the transmission risk of COVID-19 virus.}
    \label{fig:fig1}
\end{figure}

\begin{figure*}[t]
    \centering
    \includegraphics[width=0.95\linewidth]{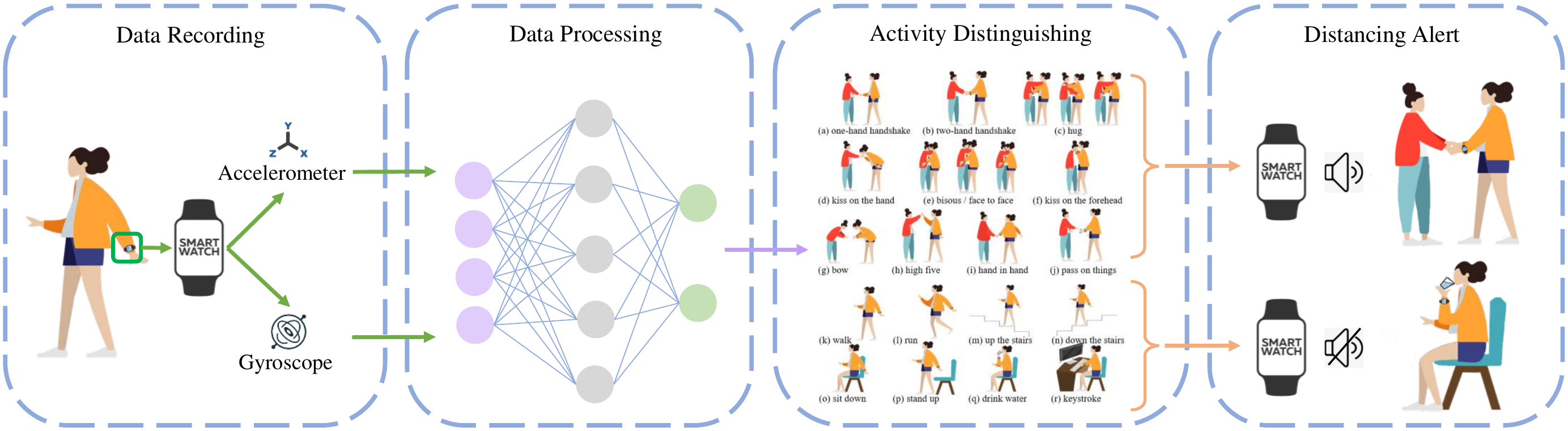}
    \caption{System Workflow. When the user wears a smartwatch, SoDA leverages the accelerometers and gyroscopes to record the wrist movements, with which SoDA conducts activity recognition and distinguishing. SoDA will alert the user to keep sufficient distance from others if it categorizes the activities to close social activities. Otherwise, SoDA keeps silent.}
    \label{fig_system}
\end{figure*}

As shown in Fig.~\ref{fig:fig1}, SoDA, equipped with smartwatches, utilizes accelerometers and gyroscopes to characterize the hand movements and recognize activities that would naturally violate the social distancing practice. It has been demonstrated that applying smartwatches to activity recognition is practical. For example, smartwatches are designed to assess the tooth brushing procedure with the Bass tooth-brushing technique~\cite{huang2016toothbrushing}, measure users' workouts~\cite{guo2017fitcoach}, and evaluate handwashing techniques in accordance with the WHO guidelines~\cite{wang2021you}. 

In this paper, we apply Vision Transformer~(ViT)~\cite{dosovitskiy2020image}, a simple yet effective deep learning model, to achieve SoDA elegantly. ViT is originally proposed for image classification and achieves impressive performance. ViT evenly divides an image into several patches as inputs and applies Transformer Encoder blocks~\cite{vaswani2017attention} on the patches. Because Transformer Encoders compute the correlation between all patches, ViT can extract features in the short-range and long-range at the very beginning, which is effective for classification. Inspired by this design, we evenly divide the time series of accelerometers and gyroscopes into several clips as inputs, conducting nothing more on the raw data. Then we feed these clips into ViT and have the classification results. If the results fall into the first 10 categories shown in Fig.~\ref{fig:fig1} (a-j), SoDA will report the social distancing practice violation to the user to promote her/his awareness of keeping sufficient social distance from others. Otherwise, if the results fall into the last 8 categories shown in Fig.~\ref{fig:fig1} (k-r), SoDA keeps silent. 

To evaluate SoDA, we recruit 10 volunteers and ask them to repeat every action shown in Fig.~\ref{fig:fig1} for 10 times, leading to a dataset with 1800 samples. We apply ViT to accomplish the action recognition task and achieve the mean accuracy of 95\%. Compared with MLP-Mixer~\cite{tolstikhin2021mlp}, ResNet~\cite{he2016deep}, and Bi-LSTM models~\cite{graves2005framewise}, ViT performs best. 
The main contributions of our work in this paper are as follows.

    
\begin{itemize}
    \item We propose SoDA to detect users' activities that may violate the social distancing practice and remind them of the violation for promoting their awareness of the social distancing practice during the COVID-19 pandemic.

    \item We collect a dataset on 10 volunteers with 1800+ samples. Extensive experimental results over the dataset show that SoDA is effective in activity recognition and social distancing alert. 
    
    \item The ablation study over five deep models demonstrates that ViT is a competing approach. We release code and envision its further use on more tasks and more modalities of time-serial data.   
\end{itemize}

\begin{figure}[t]
    \centering
    \includegraphics[width=0.98\linewidth]{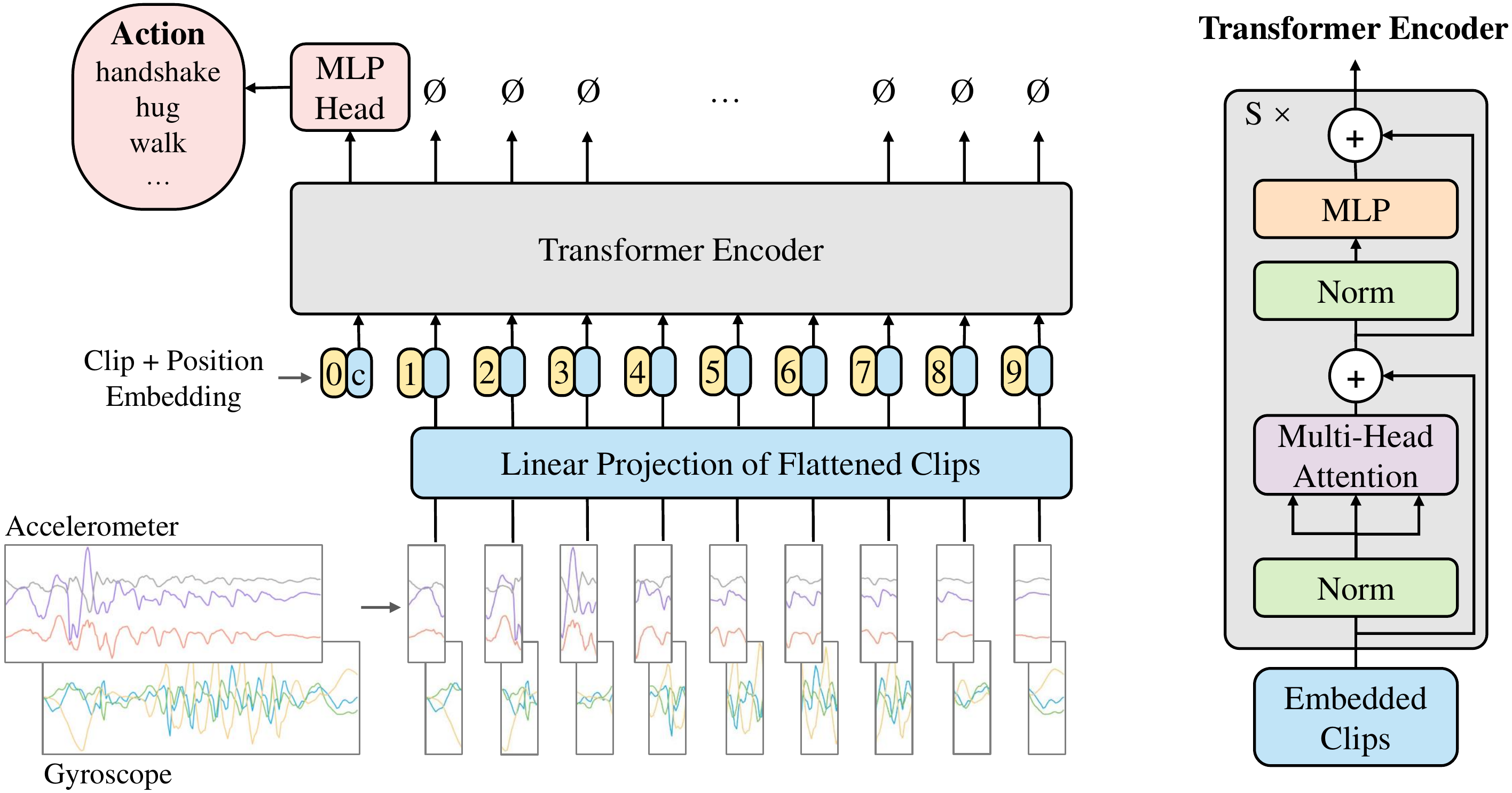}
    \caption{Model structure. Recordings from the accelerometers and gyroscopes are divided into several clips, projected into clip embedding, and fed into Transformer Encoder along with position embedding. A multi-layer perceptron head further classifies the action with the 1st output of Transformer Encoder.}
    \label{fig_net}
\end{figure}

\section{Methods}~\label{sec:methond}
\subsection{System Overview}~\label{subsec:syso}
The main workflow of SoDA is shown in Fig.~\ref{fig_system}. In data recording, accelerometers and gyroscopes continuously record wrist movements when the user wears a smartwatch. In data processing, SoDA processes the recorded time series and outputs the action category. In activity distinguishing and distancing alert, the output will be compared with eighteen activities. If it belongs to social activities, SoDA will remind the user to keep sufficient distance from others. Otherwise, if it belongs to daily activities, SoDA will keep silent.

\subsection{Deep Learning Model}~\label{subsec:deeplm}
Transformer~\cite{vaswani2017attention} is originally proposed in natural language processing (NLP). Later, Vision Transformer~(ViT)~\cite{dosovitskiy2020image} opens up the spread applications of Transformer in computer vision. In this paper, we adopt ViT in SoDA for two reasons. (1) Simple. As shown in Fig.~\ref{fig_net}, ViT is a Transformer Encoder~\cite{vaswani2017attention} plus a multi-layer perceptron head for classification. Besides, raw data of accelerometers and gyroscopes are evenly divided into several clips and fed into a linear projection directly without any pre-processing. (2) Ease on generalization. ViT has bridged the gap between NLP and computer vision. We are expecting its wide use in processing time-serial data, and take a step to demonstrate the possibility.

\textbf{\emph{Model Inputs and Linear Projection.}} Given the input data of the accelerometers and gyroscopes $x_{a} \in R^{L\times 3}$, $x_{g}\in R^{L\times 3}$, where $L$ is for the data length; 3 is the sensory dimension, we first evenly divide $x_a$ and $x_g$ into $C$ clips and reshape each clip to $R^{\frac{3L}{C}}$. Then we feed each clip into the Linear Projection to embed it with size of $R^h$. Next, we add the embedded accelerometer clip~($R^h$), embedded gyroscope clip~($R^h$), and their position embedding~(also $R^h$), to get the embedding features with the size of $R^h$. Further, we concatenate all embedding features of $C$ clips to $E \in R^{C \times h}$. At last, as shown in Fig.~\ref{fig_net}, we concatenate $E$ with a 0th-position embedding~($R^h$) and a random-clip embedding~($R^h$) for classification, and have the concatenated embedding, with size of $R^{(C+1)\times h}$, for Transformer Encoder.

\textbf{\emph{Transformer Encoder.}} Transformer Encoder is a powerful feature learner, which can be stacked in sequence. As shown in the right subfigure of Fig.~\ref{fig_net}, one Transformer Encoder is comprised of a layer normalization~(Norm)~\cite{ba2016layer}, a multi-head self-attention block~(MSA), a multi-layer perception block~(MLP), and two residual connections. MLP consists of two linear layers with a GELU non-linearity~\cite{hendrycks2016gaussian} in the middle. MSA is the key component of Transformer Encoder, described next.

\textbf{\emph{Multi-head Self-Attention~(MSA).}}~\label{sec:msa}
Recall that the inputs of Transformer Encoder are with the size of $R^{(C+1)\times h}$, after the layer normalization, the input of MSA is mapped to $U\in R^{(C+1)\times h}$. We first describe the single-head self-attention~(SSA). 

In SSA, a matrix, $W^Q \in R^{h \times h}$, maps $U$ to \textit{Query} matrix $Q$ via $UW^Q\to Q\in R^{(C+1)\times h}$.
The other two paralleled matrices, $W^K \in R^{h \times h}$, and $W^V \in R^{h \times h}$, also map $U$ to \textit{Key} matrix $K\in R^{(C+1)\times h}$ and \textit{Value} matrix $V\in R^{(C+1)\times h}$, respectively. The SSA mechanism can be written as follows.
\begin{equation}
\label{eq:ssa}
    Attention(Q,K,V,h) = softmax(\frac{QK^T}{\sqrt{h}})V
\end{equation}
where $QK^T \in R^{(C+1)\times (C+1)}$ is to compute the correlation matrix between any two clips, normalized by $\sqrt{h}$ and \textit{softmax} operator. The \textit{Value} matrix $V$ is re-weighted by the normalized correlation matrix, i.e., $softmax(\frac{QK^T}{\sqrt{h}})$, then serves as learned features of inputs. 

MSA is an advancement of SSA. In MSA, for example in $m$-head self-attention, $W^Q$ is split into $m$ smaller ones, $W^Q_i\in R^{(C+1)\times \frac{h}{m}}, i=\{1,2,...,m\}$, and maps inputs $U \to Q_i \in R^{(C+1)\times \frac{h}{m}},  i=\{1,2,...,m\}$. Similarly, $U$ is also mapped to $K_i \in R^{(C+1)\times \frac{h}{m}}$ and $V_i \in R^{(C+1)\times \frac{h}{m}}$. The MSA mechanism can be written as Equation.~\ref{eqn:msa}.
\begin{align}
\label{eqn:msa}
    &head_i = Attention(Q_i,K_i,V_i, \frac{h}{m}),  i=\{1,2,...,m\} \\
    &MultiHead(Q,K,V) = Cat(head_1,\cdots,head_m)W^O \nonumber
\end{align}
where, MSA first conducts attention mechanism over every $\{Q_i,K_i,V_i \}$ set. Then attentions from all heads are concatenated and further merged by matrix $W^O \in R^{h\times h}$. The result of MSA is $MultiHead(Q,K,V) \in R^{(C+1)\times h}$ and serves as learned features for further processes.

\textbf{\emph{Model Outputs}}.
  Transformer Encoder outputs features with the size of $(C+1)\times h$. As shown in Fig.~\ref{fig_net}, the successive MLP Head takes the first row of the outputs ($R^h$) as input to classify the activity, e.g., handshake, hug, walk, etc.   
If the result falls into the first 10 categories shown in Fig.~\ref{fig:fig1} (a-j), SoDA will remind the user to keep sufficient social distance from others. Otherwise, if the result falls into the last 8 categories shown in Fig.~\ref{fig:fig1} (k-r), SoDA keeps silent.

\subsection{Implementation Details}~\label{subsec:impde}
We construct the model with Pytorch 1.10.2 and train it with an RTX 3090 GPU. The training epoch is no more than 400. The early stop mechanism is set to save training time by stopping the training process if the model has been trained in adequate epochs and the training loss does not decrease over 40 epochs. We use AdamW~~\cite{loshchilov2017decoupled}~($\beta_1=0.9$, $\beta_2=0.999$) to optimize the model. The initial learning rate is 0.0005 and decays by a ratio of 0.5 every 40 epochs.

\section{Experiment}~\label{sec:experiment}
\subsection{Data Collection and Pre-processing}~\label{sec:data}
We recruit 10 subjects and let them wear a smartwatch branded Samsung Gear Sport on the left wrist. Subjects are required to repeat actions as shown in Fig.~\ref{fig:fig1}, where (a-f) are etiquette habits with physical contact, (g) is the etiquette habit with closed distance, (h) and (i) are daily activities with physical contact, (j) is the daily activity with indirect contact, and (k-r) are other daily activities, for 10 times. During the activity conducted, data of accelerometers and gyroscopes as well as corresponding timestamps are recorded, resulting in a dataset with 1800 samples.

In our experiments, we set the input length $L=224$ for ViT. If the length of recorded samples is longer than $224$, we will slice it into multiple segments, each with 224 sampling points, without overlapping. If the length of recorded samples or the sliced segments is within $[40, 224)$, we apply zero-padding on the samples to enlarge the length to 224. Otherwise, if the length is smaller than 40, we discard these samples. After all these processes, the dataset is with 2061 samples.

\subsection{Overall Performance}~\label{subsec:ovepe}
\textbf{(1) Cross-Validation Result.} Recall that we recruit 10 subjects to conduct 18 actions. For $i$-th action of $j$-th subject, we split the repeats into 5 groups according to the conducted time, denoted as $G_1^{i,j}, G_2^{i,j},..., G_5^{i,j}$. We apply this split strategy over all subjects and all actions, and have 5 groups of the dataset without overlap, denoted as $G_1, G_2,..., G_5$. We first utilize $\{G_2,G_3,G_4,G_5\}$ as training data to train the model of ViT-MS/8~(see Table~\ref{tab:par_four}), leaving $G_1$ to test the trained model. 
As listed in Table~\ref{tab:folds}, the classification accuracy over $G_1$ is 0.914~(91.4\%). We then utilize $\{G_3,G_4,G_5,G_1\}$ as training data, leaving $G_2$ for testing. Similarly, we repeat this training-then-test as listed in Table~\ref{tab:folds} and have the mean accuracy of 0.947~(94.7\%) over this 5-fold cross-validation. 

\begin{table}[h]
\centering
\caption{Five-fold cross-validation accuracy of ViT-MS/8.}
\label{tab:folds}
\renewcommand{\arraystretch}{1.2}
\begin{tabular}{ccccccc}
\hline
Test Group &$G_1$  & $G_2$  & $G_3$  & $G_4$  & $G_5$  & Mean \\ \hline
Accuracy & 0.914   & 0.956   & 0.970   & 0.976   & 0.920   & 0.947 \\ \hline
\end{tabular}
\end{table}

\begin{table*}[t]
\centering
\caption{Parameters and Accuracy of ViTs, MLP-Mixers, ResNets, and Bi-LSTMs.}
\label{tab:par_four}
\renewcommand{\arraystretch}{1.2}
\begin{tabular}{ccccccccccc}
\hline
Model & Blocks & Hidden Dim & Heads & Clip Length & Parameters(M) & FLOPs(M) & Accuracy & Precision & Recall & F1  \\ \hline
\multicolumn{11}{c}{ViT}                                          \\ \hline
ES/8  & 2      & 128        & 4     & 8          & 0.43         & 12.10     & 0.942 & 0.945 & 0.942 & 0.942 \\
ES/16 & 2      & 128        & 4     & 16         & 0.44         & 6.23      & 0.924 & 0.929 & 0.923 & 0.924 \\
ES/32 & 2      & 128        & 4     & 32         & 0.45         & 3.38      & 0.885 & 0.887 & 0.884 & 0.883  \\
MS/8  & 4      & 256        & 4     & 8          & 3.23         & 93.63     & \textbf{0.947} & \textbf{0.948} & \textbf{0.948} & \textbf{0.947} \\
MS/16 & 4      & 256        & 4     & 16         & 3.24         & 48.16     & 0.935 & 0.938 & 0.934 & 0.934 \\
MS/32 & 4      & 256        & 4     & 32         & 3.27         & 25.73     & 0.904 & 0.909 & 0.903 & 0.903 \\
S/8   & 8      & 512        & 8     & 8          & 25.36        & 739       & 0.926 & 0.929 & 0.926 & 0.926    \\
S/16  & 8      & 512        & 8     & 16         & 25.38        & 381       & 0.923 & 0.924 & 0.922 & 0.920    \\
S/32  & 8      & 512        & 8     & 32         & 25.43        & 203       & 0.899 & 0.902 & 0.897 & 0.897    \\ \hline
\multicolumn{11}{c}{MLP-Mixer~\cite{tolstikhin2021mlp}}                                    \\ \hline
ES/8  & 2      & 128        & $\backslash$  & 8     & 0.32  &  21.45    & 0.910 & 0.916 & 0.909 & 0.910 \\
ES/16 & 2      & 128        & $\backslash$  & 16    & 0.29  &  9.21     & 0.909 & 0.914 & 0.910 & 0.909 \\
ES/32 & 2      & 128        & $\backslash$  & 32    & 0.29  &  4.29     & 0.886 & 0.891 & 0.884 & 0.885 \\
MS/8  & 4      & 256        & $\backslash$  & 8     & 2.22  &  144      & 0.912 & 0.916 & 0.913 & 0.912 \\
MS/16 & 4      & 256        & $\backslash$  & 16    & 2.16  &  65.81    & 0.910 & 0.913 & 0.910 & 0.909 \\
MS/32 & 4      & 256        & $\backslash$  & 32    & 2.16  &  31.47    & 0.897 & 0.903 & 0.896 & 0.896 \\
S/8   & 8      & 512        & $\backslash$  & 8     & 17.04 &  1045     & 0.895 & 0.897 & 0.895 & 0.894 \\
S/16  & 8      & 512        & $\backslash$  & 16    & 16.91 &  497      & 0.898 & 0.901 & 0.897 & 0.897 \\
S/32  & 8      & 512        & $\backslash$  & 32    & 16.92 &  243      & 0.899 & 0.902 & 0.897 & 0.897 \\ \hline
\multicolumn{11}{c}{ResNet~\cite{he2016deep}}                                      \\ \hline
Res18    & $\backslash$      & 256    & $\backslash$  & $\backslash$  & 3.85      &  38.75    & 0.920 & 0.924 & 0.918 & 0.919    \\
Res34    & $\backslash$      & 256    & $\backslash$  & $\backslash$  & 7.22      &  78.95    & 0.891 & 0.897 & 0.891 & 0.892    \\
Res50    & $\backslash$      & 512    & $\backslash$  & $\backslash$  & 15.96     &  177      & 0.869 & 0.874 & 0.867 & 0.868    \\
Res101   & $\backslash$     & 512    & $\backslash$  & $\backslash$  & 28.26     &  351      & 0.848 & 0.854 & 0.850 & 0.849    \\ \hline
\multicolumn{11}{c}{Bi-LSTM~\cite{graves2005framewise}}                                      \\ \hline
ES    & $\backslash$      & 128        & $\backslash$  & $\backslash$  & 0.27      &  0.70     & 0.853 & 0.853 & 0.851 & 0.849   \\
MS    & $\backslash$      & 256        & $\backslash$  & $\backslash$  & 2.63      &  4.54     & 0.935 & 0.938 & 0.935 & 0.935   \\
S     & $\backslash$      & 512        & $\backslash$  & $\backslash$  & 16.81     &  25.85    & 0.936 & 0.938 & 0.936 & 0.935   \\ \hline
\end{tabular}
\end{table*}

\begin{figure}[t]
    \centering
    \includegraphics[width=1\linewidth]{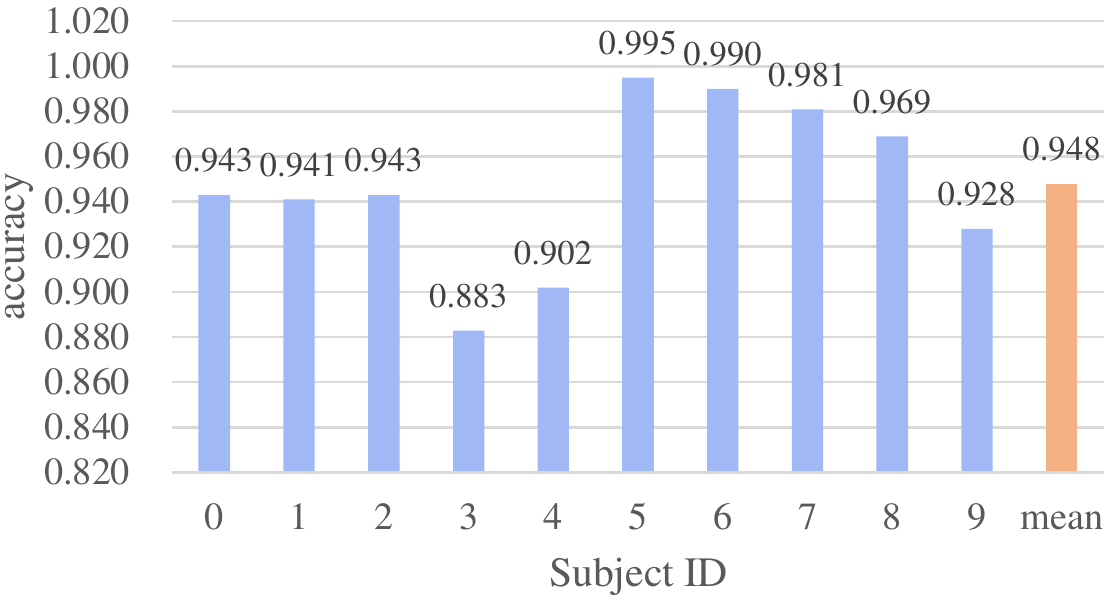}
    \caption{SoDA with ViT-MS/8 performs well over all the subjects. The mean accuracy over all subjects is 0.948. }
    \label{fig:overall_subject}
\end{figure}

\textbf{(2) Accuracy over subjects.}
We show the mean accuracy of the cross-validation procedure in another fine-grained view, i.e., accuracy over subjects. Denote the accuracy of the $i$-th activity as $a_i$, which can be computed as follows.
\begin{equation}\label{eq:acc_actvity}
    a_i = \frac{\sum_{j=1}^{\left| S_{}i\right|}  \left\| p_{i}^{j},gt_{i}^{j}\right\|  }{\left| S_i\right|}
\end{equation}
where $\left| S_i\right|$ is for the sample number of the $i$-th subjects; $gt_{i}^{j}$ and $p_i^j$ are for the ground-truth and prediction of the $j$-th sample of the $i$-th subject; $\left\| p_{i}^{j},gt_{i}^{j}\right\| $ returns 1 if $p_i^j$ equals $gt_{i}^{j}$, otherwise, returns 0.

\begin{figure}[t]
    \centering
    \includegraphics[width=1\linewidth]{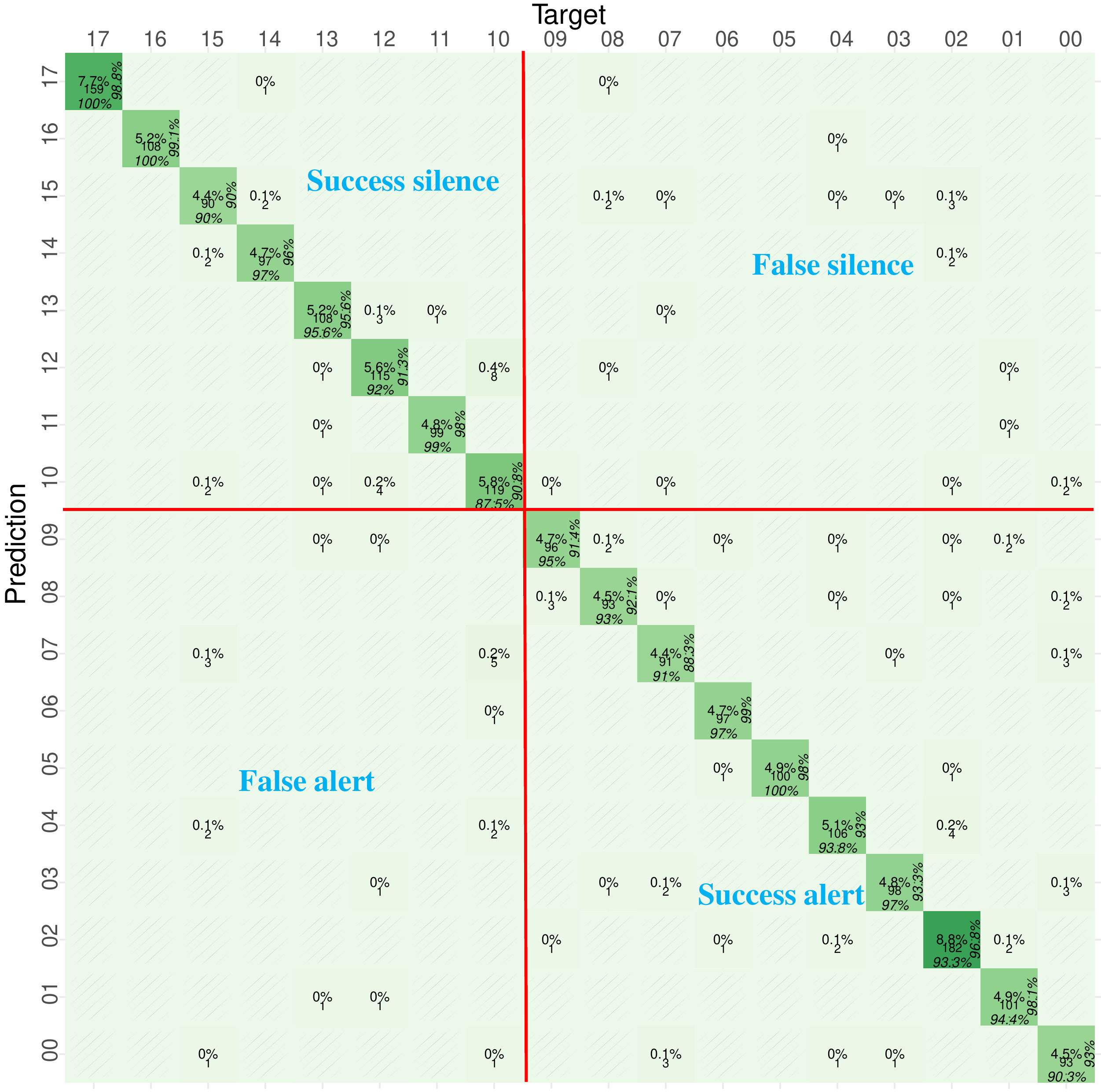}
    \caption{Confusion matrix over 18 activities. SoDA classifies all activities well. Besides, precisions of success alert~(alert when should alert), success silence~(silence when should keep silence), false alert~(should alert but keeps silent), false silence~(should be silent but alert) are 0.982, 0.978, 0.018 ,and 0.022 respectively,  demonstrating that SoDA can conduct social distancing alert precisely.}
    \label{fig:confusion}
\end{figure}

As Fig.~\ref{fig:overall_subject} shows, SoDA with ViT-MS/8 performs well over all subjects, especially over the 5th~(0.995), 6th~(0.990), 7th~(0.981), and 8th~(0.969) subjects. The mean accuracy over all subjects is 0.948.

\textbf{(3) Performance over activities.}
We further show the confusion matrix over activities in Fig.~\ref{fig:confusion}, along with the number of samples, precision, and recall. As the figure shows, SoDA achieves good precision and recall over all the activities, especially 
over the 5th~(kiss on the forehead), 6th~(bow), 16th~(drink water), and 17th~(keystroke). One most error happens in the classification of the 0th activity~(one-hand shake). This is because the smartwatch is not worn on this shaking hand. However, this error would not lead to much false silence~(should alert but keep silent). As shown in Fig.~\ref{fig:confusion}, most false silence happens at recognition on the 2nd activity~(hug). Meanwhile, the most false alert~(should be silent but alert) happens at recognition on the 10th activity~(walk). Besides, precisions of success alert, success silence, false alert, and false silence are 0.982, 0.978, 0.018, and 0.022, respectively.

\textbf{(4) Ablation study on ViT variants.}

\begin{figure}[t]
    \centering
    \includegraphics[width=1\linewidth]{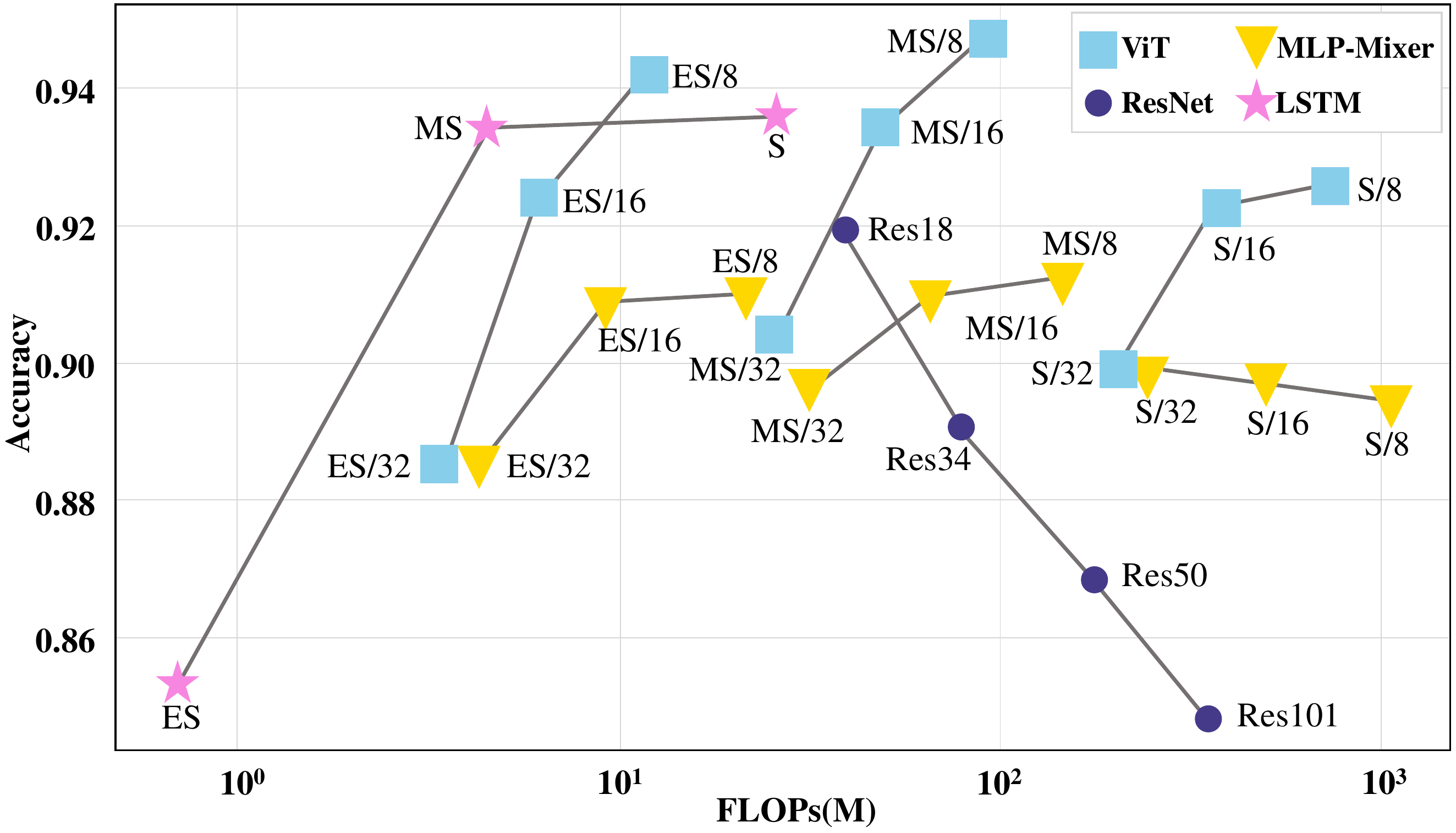}
    \caption{FLOPs and accuracy over ViT, MLP-Mixer, ResNet, and Bi-LSTM. The results show that ViT models achieve best with similar FLOPs. All large models face the problem of over-fitting. Bi-LSTM is a competing method, however, training Bi-LSTM-S costs $\times11.3$ times than ViT-MS/32.}
    \label{fig_accuracy}
\end{figure}

Model scale. The Model scale depends on the number of Transformer Encoder blocks, hidden dimensions, self-attention heads, etc. As listed in Table~\ref{tab:par_four}, we set the ViT model in 3 scales, i.e., ES~(extra small), MS~(medium small), and S~(small).
As shown in Fig.~\ref{fig_accuracy}, when we expand the scale from ES to MS, the accuracy increases. However, the accuracy decreases when we expand the scale to S, which indicates overfitting happens.

Clip length. In each scale, the accuracy decreases when we lengthen the clip length, as shown in Table \ref{tab:par_four}. This indicates that decreasing the clip length (increasing clip number) could result in a better representation to achieve better accuracy. However, better accuracy is traded with more FLOPs. In the network design practice, the trade-off between accuracy and FLOPs deserves to be evaluated.

\begin{figure}[t]
    \centering
    \includegraphics[width=1\linewidth]{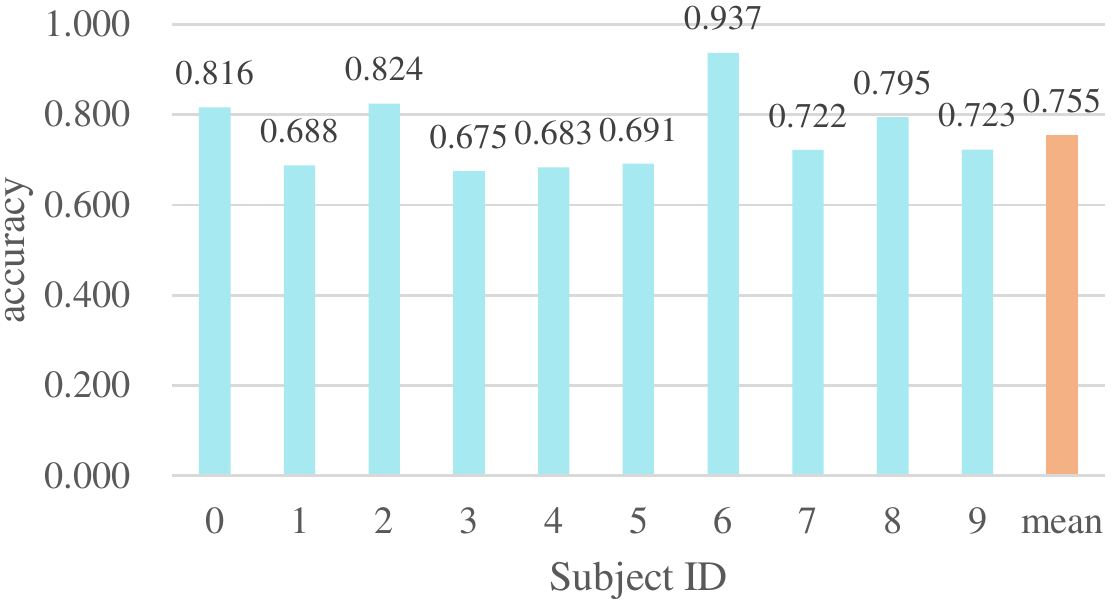}
    \caption{Accuracy of SoDA with ViT-MS/8 on unseen subjects in the leave-one-subject-out evaluation manner. The accuracy is around 0.7-0.8, and the mean accuracy is 0.76.}
    \label{fig:unseen}
\end{figure}

\subsection{Comparison with Different Networks}~\label{sec:compare}
We compare the results of ViT models with the MLP-Mixer~\cite{tolstikhin2021mlp}, ResNet~\cite{he2016deep}, and Bi-LSTM~\cite{graves2005framewise} models. The specification and accuracy of these models are listed in Table~\ref{tab:par_four}. To facilitate the understanding of these values, we show the accuracy and FLOPs in Fig.~\ref{fig_accuracy}. The figure clearly shows that (1) ViT models work better than other models with similar FLOPs, and ViT-MS/8 model performs best. (2) Larger models are not always better, e.g., ViT-S, ResNet101, MLP-Mixer-S, which poses the over-fitting problem for large models on our dataset. (3) Compared with ViT-MS/32, Bi-LSTM-S is a competing approach with similar FLOPs. However, training Bi-LSTM-S costs approximately $\times 11.3$ times longer than ViT-MS/32. Besides, its inferring time is also $\times 2.8$ times longer.

\begin{table}[t]
\caption{Accuracy over unseen subjects.}
\centering
\renewcommand{\arraystretch}{1.2}
\begin{tabular}{cccc}
\hline
Model    & ViT-ES/8 & ViT-MS/8 & ViT-S/8 \\ \hline
Accuracy & 0.779     & 0.755     & 0.739    \\ \hline
\end{tabular}
\label{tab:unseen}
\end{table}

\subsection{Accuracy over unseen Subjects}
To evaluate SoDA on unseen subjects, we apply the leave-one-subject-out~(LOSO) manner. That is, for example, we first use the data of the 2nd-10th subjects to train a ViT-MS/8, and use the data of the 1st subject to testing the trained model. Then we use the data of the 3nd-10th and 1st subjects to train another ViT-MS/8, and use the data of the 2nd subject to testing the trained model. We conduct this LOSO for all subjects in sequence and have 10 trained ViT models to be evaluated. The evaluation is reported in Fig.~\ref{fig:unseen}, which shows that the accuracies over the unseen subject are around 0.7-0.8, and the mean accuracy is 0.755. Further, we apply ViT-ES/8 and ViT-S/8 to the LOSO evaluation and list the accuracy in Table~\ref{tab:unseen}. The table shows that the best performance over unseen subjects is achieved by the smallest ViT-ES/8, i.e., 0.779. Promoting the performance over unseen subjects should be valuable future work, thus we would like to release a dataset to facilitate research on this topic.

\section{Related Work}~\label{sec:related_work}
\subsection{Social Distance Detection Methods}~\label{sec:sddm}
In social distance detection, cameras are commonly used to collect videos of the monitored areas~\cite{ghasemi2021auto}. They measure the distance or density of people in videos and evaluate whether their social distance complies with regulations. It is a very intuitive method. However, it may have the effect of viewing angle occlusion, so that a part of the blocked people cannot be detected correctly. And cameras installed in private places may expose personal privacy. Positioning technologies, such as Bluetooth Low Energy (BLE)~\cite{chandel2020proxitrak}, WiFi~\cite{kanjo2021crowdtracing}, and Ultra-Wide Band (UWB)~\cite{reddy2020social}, are also adopted to detect social distance. But most of these methods tend to evaluate indoor activities. 
Various sensors are also exploited to estimate the social distance of people, such as thermal sensors~\cite{naser2021novel}, vibration sensors~\cite{dong2021social}, and magnetic field sensors~\cite{bian2020wearable}.


\subsection{Applications of Smartwatch in Health}~\label{subsec:aosih}
As a wearable device equipped with multiple sensors, the smartwatch has many applications in the field of healthcare. Combined with deep learning algorithms, it can use data from accelerometer and gyroscope to perform action recognition, e.g. detecting falls of the elderly~\cite{csengul2022deep}, in this paper, the author adopts the bi-directional long short term memory~(Bi-LSTM) neural network to classify the fall detection from the common daily activities. Besides, smartwatches have a variety of applications in sleep, such as capturing a range of information about sleep quality~\cite{chang2018sleepguard}, detecting early Parkinson's disease through sleep~\cite{iakovakis2020smartwatch}, detecting sleep apnea~\cite{chen2021apneadetector}, and monitoring breathing rate and body movement during sleep~\cite{sun2017sleepmonitor}. These studies enable people to obtain a lot of health-related information from sleep merely with a single smartwatch. 


\section{Conclusion}~\label{sec:conclusion}
In this paper, we present SoDA, a smartwatch-based solution, to detect users' activities that may violate the social distancing practice, reminding people of the practice during the current COVID-19 pandemic to prevent virus transmission. To evaluate SoDA, we recruit 10 volunteers and build a dataset with 1800+ samples. Experimental results show that SoDA with simple ViT models is efficient to distinguish 10 social activities from daily activities with promising performance. Deserve to mention that we show ViT generalizes well from handling image data to the accelerometer and gyroscope data, with the great potentiality to generalize to more modalities of time-serial data.

{\small
\bibliographystyle{./bibliography/IEEEtran}
\bibliography{./bibliography/reference}
}

\end{document}